\documentclass[30,preprint]{elsarticle}
\usepackage{amsmath,amssymb,graphicx,booktabs,balance}
\usepackage[hidelinks]{hyperref}
\journal{Physics Letters B}
\newcommand{\Tr}{\operatorname{Tr}}
\newcommand{\Herm}{\operatorname{Herm}}
\newcommand{\id}{\mathbf{1}}
\begin{document}
\begin{frontmatter}
\title{Holographic $p$-wave magnetism: finite-wavevector order and odd-parity electron spectra}
\author{Wen-Yu Wen\corref{cor}}
\ead{wenw@cycu.edu.tw}
\cortext[cor]{Corresponding author.}
\address{Department of Physics and Center for High Energy Physics, Chung Yuan Christian University, Chung Li City, Taiwan}
\begin{abstract}
We construct a finite-density gauge/gravity model in which a neutral spin triplet orders at nonzero wavevector and generates an odd-parity electronic spin response. A momentum-dependent double-trace deformation favors finite-wavevector order, while a probe Dirac doublet determines the retarded self-energy of a hybridized electron band. For one set of couplings, the source-free circular branch emerges at $T_N/\mu=0.2880$ and $Q_N/\mu=0.8767$ and has lower grand potential than the normal state at all sampled temperatures. The longitudinal laboratory-frame spectra obey $A_s(\omega,k)=A_{-s}(\omega,-k)$, even in the presence of dissipation. Near onset, numerical exponents of $0.4997$, $0.5008$ and $1.0001$ for the condensate, static exchange and representative spin polarization confirm distinct square-root and linear onsets, the latter consistent with vector chirality. At $T=0.649T_N$, the Hermitian spin-mixing scale at zero generalized momentum changes from $0.0445\mu$ to $0.0707\mu$ between $\omega/\mu=-0.2$ and $0.2$; dispersive ridges persist across the Fermi level. A separate exact lattice benchmark explains the coexistence of reconstruction, a spin node and gapless bands. The result links spontaneous finite-wavevector order to a dynamical odd-parity electron self-energy within a probe-limit model. It captures qualitative features of EuSi, but neither establishes global magnetic stability nor reproduces its material-specific reconstruction energies or transport.
\end{abstract}
\end{frontmatter}
% Avoid stretching paragraph glue on the first page.
\raggedbottom
\AddToHookNext{shipout/after}{\flushbottom}

\section{Introduction}
Odd-parity magnetic spin textures offer a route to momentum-dependent electron spin polarization without uniform magnetization or explicit spin--orbit coupling. In a circular coplanar spiral, electrons with opposite spin projections perpendicular to the spiral plane scatter into different intermediate momenta. The spectral spin difference can therefore be odd in momentum and integrate to zero. This is a characteristic electronic signature of nonrelativistic $p$-wave magnets \cite{hellenes,brekke,yu}. Spin-triplet Pomeranchuk order provides an earlier route to unconventional spin textures \cite{wuspin}, but conservation laws constrain instabilities of homogeneous current operators \cite{kiselev,wuklein}. Here the primary order is a finite-wavevector magnetic condensate, not a homogeneous $l=1$ spin-current order parameter. Recent examples include electrical control in a spiral multiferroic and a metallic magnet with a commensurate spin helix \cite{song,yamada}.

In EuSi, neutron diffraction identifies a coplanar spiral between 33.4 and 62 K, with propagation vector $(0,1/3,0)$ in crystallographic reciprocal units. The magnetic structure preserves time reversal combined with a lattice translation. Spin-resolved photoemission reveals a sign-changing $S_y$ signal across magnetic Brillouin zones and suppression near nodal regions, together with substantial band reconstruction \cite{cheng}. The relevant comparison is with spin-resolved spectral weight at physical momentum, rather than eigenvalues alone.

Generalized Bloch theory provides a description of spiral electronic states \cite{sandratskii,larsen}, and recent studies address chirality-dependent spin textures \cite{gobel} and the tendency toward incommensurate order \cite{lee}. A prescribed static exchange alone does not determine the equilibrium pitch or the dynamical self-energy of an electron coupled to a dissipative sector. Gauge/gravity duality provides a setting in which an order parameter and a retarded response can be computed from bulk boundary-value problems. Holographic magnetic order, striped instabilities and helical current phases have been studied in Refs.~\cite{iqbal,stripes,donos}; fermionic spectra in spatially modulated holographic backgrounds have also been investigated \cite{bianchi}. Our construction differs in its use of an internal spin rotation, a competing-exchange double-trace deformation and a semiholographic electron coupled to the same ordered magnetic saddle. The new calculation ties the equilibrium pitch to the dissipative spectral response and tests the distinct critical onsets; neither helical order nor odd-parity spin splitting is new by itself.

Our primary objective is to construct a source-free circular magnetic branch at finite density and calculate its preferred wavevector and spin-dependent retarded electron self-energy within one model. The secondary objectives are to establish the laboratory-frame parity relation independently of numerical fits, derive and test the critical scaling of magnetic mixing and spin polarization, distinguish magnetic mixing from spectral peak splitting, and identify which aspects of EuSi can and cannot be addressed by this continuum construction.

\section{Finite-wavevector magnetic order}
At fixed electric chemical potential $\mu$, the five-dimensional bosonic action is
\begin{align}
S_{\rm bulk}&=\frac{1}{2\kappa_5^2}\int d^5x\sqrt{-g}\left(\mathcal L_{\rm EM}-\lambda_N\mathcal L_N\right),\nonumber\\
\mathcal L_{\rm EM}&=R+\frac{12}{L^2}-\frac{Z_c}{4}F_{MN}F^{MN},\nonumber\\
\mathcal L_N&=\frac{Z_s}{4}\mathcal F^a_{MN}\mathcal F^{aMN}
+\frac12(D_M\Phi)^a(D^M\Phi)^a+V,\nonumber\\
V&=\frac{m^2}{2}\Phi^2+\frac{u}{4}(\Phi^2)^2,
\qquad \Phi^2=\Phi^a\Phi^a,\quad u>0.\label{eq:action}
\end{align}
Here $F=dA$ is the electric $U(1)$ field strength, while $a_M^a$ is an independent $SU(2)$ connection with $\mathcal F^a_{MN}=\partial_M a_N^a-\partial_N a_M^a+\epsilon^{abc}a_M^b a_N^c$ and $(D_M\Phi)^a=\partial_M\Phi^a+\epsilon^{abc}a_M^b\Phi^c$. The electrically neutral triplet is dual to a magnetic order parameter. We set $L=Z_c=1$, $Z_s=1/g_s^2$, $m^2=-15/4$ and $u=g_s^2=1$. The coefficient $\lambda_N=2\kappa_5^2 C_N$ controls the magnetic stress tensor. We first vary the spin fields at finite $\lambda_N$ and then neglect their stress tensor in the Einstein equations ($\lambda_N\to0$). The magnetic grand-potential difference is quoted per $C_N$; the order parameter is given in its chosen probe normalization. The electron does not feed back on the magnetic saddle.

Unless explicit units are restored, boundary variables are measured in powers of $\mu$ ($\mu=1$). The dimensionless radial coordinate $z$ runs from the AdS boundary at 0 to the horizon at 1; $z_h=\sqrt{3}b$ is the horizon length in units of $\mu^{-1}$, with $0<b<\sqrt{2}$. The RN--AdS$_5$ background is
\begin{align}
ds^2&=z^{-2}[-f\,d\hat t^2+dz^2/f+d\hat{\boldsymbol{x}}^2],\nonumber\\
f&=1-(1+b^2)z^4+b^2z^6,\label{eq:rn}\\
\hat A_t&=z_h(1-z^2),\qquad T/\mu=\frac{2-b^2}{2\pi\sqrt{3}b}.
\label{eq:temp}
\end{align}
Hatted boundary coordinates are rescaled by $z_h$. Choose $\boldsymbol{\Phi}=\varphi(z)(\cos(\hat q\hat y),0,\sin(\hat q\hat y))$ and $a^y_{\hat y}=a(z)$, where the superscript denotes the internal spin axis and $\hat q=z_h Q$. Then the covariant gradient has magnitude $(\hat q-a)\varphi$. The radial equations are
\begin{align}
(fz^{-3}\varphi')'&=z^{-3}(\hat q-a)^2\varphi+z^{-5}(m^2\varphi+u\varphi^3),\label{eq:phi}\\
(fz^{-1}a')'&=-g_s^2z^{-3}(\hat q-a)\varphi^2.
\label{eq:a}
\end{align}
The scalar exponents $\Delta_-=3/2$ and $\Delta_+=5/2$ lie in the AdS$_5$ alternative-quantization window \cite{klebanov}. Write $\varphi=\alpha z^{3/2}+\beta z^{5/2}+\cdots$ and define $N/\mu^{3/2}=\alpha/z_h^{3/2}$. A double-trace deformation \cite{witten,multitrace} implements exchange competition through
\begin{align}
a(0)&=0,\quad J_N=\beta-z_h\kappa(Q)\alpha=0,\nonumber\\
\kappa(Q)&=0.05-2Q^2+Q^4.\label{eq:bc}
\end{align}
Here $J_N$ is proportional (up to sign and scale) to the deformed scalar source; the physical kernel has dimension one, $\kappa_{\rm phys}(Q_{\rm phys})=\mu\kappa(Q_{\rm phys}/\mu)$. The double-trace interaction is quadratic in the magnetic operator and even under $Q\to-Q$. It explicitly selects the propagation axis $y$ and favors $|Q|=1$ before the bulk gradient cost is included. Thus the finite-wavevector preference is engineered by exchange competition, rather than arising intrinsically from the undeformed RN model. Unlike a linear magnetic source, the deformation does not impose a nonzero condensate: $J_N=0$ permits $\alpha\neq0$ and $\beta\neq0$. The amplitude develops spontaneously, while the equilibrium pitch is determined by the full magnetic free energy. For a fixed spin frame, opposite senses of rotation are degenerate under $(Q,a)\to(-Q,-a)$. They are also related by a global spin rotation by $\pi$ about the spin $x$ axis; without spin anisotropy they do not constitute independent Ising-like chiral sectors.

In the normal state ($\varphi=a=0$), a regular linear scalar solution yields the inverse susceptibility in our normalization, $r(T,Q)=\kappa(Q)-\beta/(z_h\alpha)$. The onset on the sampled interval $0\leq Q\leq1.4$ follows from $r=\partial_Qr=0$ at a local minimum of $r$. We obtain $T_N/\mu=0.287998$, $Q_N/\mu=0.876744$ and $r(T_N,0)=0.54436$. An independent shooting check of the linear problem gives $r_{QQ}\simeq5.74>0$ at onset (Appendix~\ref{app:num}). This identifies a finite-$Q$ instability in the channel considered, not a proof of the globally leading instability among all saddles.

Below onset we solve Eqs.~\eqref{eq:phi}--\eqref{eq:bc} and the stationarity condition at fixed $T,\mu$ (Appendix~\ref{app:num}):
\begin{equation}
\int_0^1 dz\,\frac{(\hat q-a)\varphi^2}{z^3}+\frac12\kappa'(Q)\alpha^2=0.
\label{eq:pitch}
\end{equation}
\begin{figure}[t]
\centering\includegraphics[width=\columnwidth]{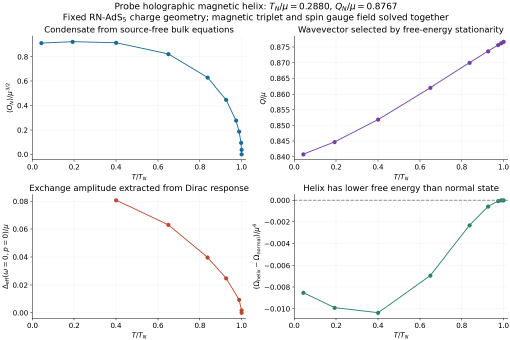}
\caption{Magnetic and electronic scales for the circular branch. Top: condensate and stationary pitch from Eqs.~\eqref{eq:phi}--\eqref{eq:pitch}. Bottom: static exchange from Eq.~\eqref{eq:delta} (computed at a subset of temperatures) and normalized grand-potential difference relative to the normal state at fixed $T,\mu$. Lines connect sampled points; they do not represent a fit or extrapolation.}
\label{fig:mag}
\end{figure}
Figure~\ref{fig:mag} shows a continuous onset. The extended near-critical calculation below gives a condensate exponent of $0.4997$, consistent with mean-field behavior. Fit-window variation rather than statistical confidence intervals characterizes this deterministic calculation. The ordered solution has lower grand potential than the normal state at all sampled temperatures. At $T/T_N=0.6493$, constrained solutions at $Q=0.85206,0.86206,0.87206$ give $\delta\Omega/\mu^4=-0.0068444,-0.0069327,-0.0068401$, respectively. The centered finite-difference curvature is $1.81$ in scaled units. This tests local stability with respect to pitch variations on the circular branch only. Neither the relative free energies of collinear, conical or multi-$Q$ phases nor the full fluctuation spectrum have been computed.

\section{Retarded electron response}
We couple a spin-independent elementary electron band to a fermionic operator in the magnetic sector using the semiholographic prescription \cite{faulknerpolchinski}. The bulk spinor is massless, has electric charge $q_\Psi=1$, and transforms as a fundamental doublet of the internal $SU(2)$, with physical spin matrices $\tau_a$. Its minimally coupled Dirac operator uses $\mathcal D_M=\nabla_M-iq_\Psi A_M-ia_M^a\tau_a/2$ and a Yukawa mass $y_\Psi\Phi^a\tau_a$, with $y_\Psi=1$. The internal rotation $R(y)=\exp(iQy\tau_y/2)$ removes the explicit spatial dependence of this mass, the continuum analogue of the generalized Bloch construction \cite{sandratskii,larsen}. On the longitudinal cut (transverse momenta zero), the rotating-frame matrices are
\begin{equation}
M=y_\Psi\varphi\tau_x,\qquad K=z_hp\id+\tfrac12(\hat q-a)\tau_y.
\label{eq:mk}
\end{equation}
We use the real-time holographic prescription \cite{iqballiu,faulkner}. The ingoing Dirac equations and UV Green-function convention are given in Appendix~\ref{app:num}. The two spacetime helicities are averaged with equal weight to define the spin response; this is the specified operator projection, not a material orbital matrix element. The retarded self-energy is $\Sigma^R=h^2G^R$, with $h^2/\mu=0.10$ fixed across the transition. In this massless AdS$_5$ convention $G^R$ is dimensionless. The elementary dispersion is $\epsilon(k)/\mu=0.3(k/\mu)^2-0.3$ when dimensions are restored.

Let $P_s=(\id+s\tau_y)/2$ ($s=\pm1$). In the rotating frame the electron inverse propagator is
\begin{equation}
G_c^{-1}(\omega,p)=(\omega+i0^+)\id-\sum_s P_s\epsilon(p+sQ/2)-\Sigma^R(\omega,p).
\label{eq:dyson}
\end{equation}
The spin-$s$ component of $R(y)e^{ipy}$ carries laboratory momentum $k=p+sQ/2$. Accordingly, the spin-resolved spectra at a common physical momentum are
\begin{align}
g_s(\omega,k)&=\Tr[P_s G_c(\omega,k-sQ/2)],\nonumber\\
A_s&=-\pi^{-1}\Im g_s,\qquad P_y=\frac{A_+-A_-}{A_++A_-}.
\label{eq:spectral}
\end{align}
The polarization is defined only where the denominator is nonzero. In particular, comparing the two diagonal matrix elements at the same $p$ would generally compare different physical momenta.

The Schur complement gives the effective diagonal laboratory-frame self-energy
\begin{align}
\Sigma^R_{{\rm eff},s}&=\omega+i0^+-\epsilon(k)-g_s^{-1}\nonumber\\
&=\Sigma^R_{ss}+\frac{\Sigma^R_{s,-s}\Sigma^R_{-s,s}}
{\omega+i0^+-\epsilon(k-sQ)-\Sigma^R_{-s,-s}}.
\label{eq:schur}
\end{align}
All matrix elements on the right are evaluated at $(\omega,p=k-sQ/2)$. The numerator is a product of two retarded amplitudes, not a modulus squared. A useful static exchange diagnostic is
\begin{equation}
\Delta_{\rm eff}(T)=|[\Herm\Sigma^R(0,0;T)]_{+,-}|,
\qquad \Herm X=(X+X^\dagger)/2.
\label{eq:delta}
\end{equation}
Here $p=0$ corresponds to bare partner momenta $\pm Q/2$. Their common bare energy need not be zero. Consequently, $2\Delta_{\rm eff}$ is not in general an observable peak separation: diagonal self-energies, frequency dispersion and damping enter the pole condition $\det G_c^{-1}=0$ and the spectral maxima.

\subsection{Exact parity and near-onset behavior}
Conjugation by $\tau_x$ preserves $M$ and sends $K(p)$ to $-K(-p)$. Exchanging the two spacetime helicities in the radial equations restores the ingoing problem. With equal helicity weights, $G^R(\omega,-p)=\tau_x G^R(\omega,p)\tau_x$. The even bare band and $\tau_xP_s\tau_x=P_{-s}$ imply
\begin{equation}
A_s(\omega,k)=A_{-s}(\omega,-k),\quad P_y(\omega,-k)=-P_y(\omega,k).
\label{eq:odd}
\end{equation}
This identity is exact in the longitudinal model and does not require a real or frequency-independent self-energy. It forces the spin difference to vanish at $k=0$ even when a reconstructed peak remains. It also implies zero integrated longitudinal spin difference on a symmetric momentum interval. The full crystal's nodal planes are not established by this calculation.

A further consequence concerns the approach to onset. The gauge equation gives $a=O(N^2)$, while $M=O(N)$. Reversing the amplitude is equivalent to conjugation by $\tau_y$, which commutes with $K$. For a smooth expansion about a regular normal-state solution, the diagonal self-energy corrections are even in $N$, whereas off-diagonal elements are odd. Thus, at fixed nonsingular $(\omega,k)$,
\begin{equation}
\Delta_{\rm eff}=O(|N|),\qquad A_+-A_-=O(N^2).
\label{eq:scaling}
\end{equation}
The normal spin response is zero in laboratory momentum for any choice of rotating-frame pitch. Equation~\eqref{eq:scaling} therefore also applies along a smooth circular branch. If the linear mixing coefficient is nonzero and the spectral denominator remains finite, mean-field onset predicts $\Delta_{\rm eff}\propto (1-T/T_N)^{1/2}$ and $P_y=O(1-T/T_N)$ at such fixed points. The expansion also requires a nonvanishing quadratic polarization coefficient for a linear onset; symmetry-enforced nodes are excluded. It is not uniform near a degenerate sharp pole, where level splitting can be linear in $|N|$.

The quadratic polarization onset admits a vector-chirality interpretation. Write the boundary texture as $\boldsymbol N(y)=N(\cos Qy,0,\sin Qy)$, in chemical-potential units. Its vector chirality is
\begin{equation}
\boldsymbol{\mathcal C}=\boldsymbol N\times\partial_y\boldsymbol N
=-QN^2\hat{\boldsymbol y}_{\rm spin}.
\label{eq:chirality}
\end{equation}
It is even under amplitude reversal but odd under reversal of the spiral at fixed spin axes. In Eq.~\eqref{eq:mk}, reversing $(Q,a)$ at fixed $p$ is equivalent to conjugation by $\tau_x$. The projectors and momentum shifts in Eq.~\eqref{eq:spectral} then imply $P_y(\omega,k;-Q)=-P_y(\omega,k;Q)$. Together with Eq.~\eqref{eq:odd} and amplitude parity, this gives, near onset and for small $k$ at fixed nonzero $Q$,
\begin{equation}
P_y(\omega,k)=c(\omega,T,Q^2)\,kQN^2
+O(k^3N^2,kN^4).
\label{eq:chiralitypol}
\end{equation}
The coefficient $c$ depends on the full electronic response; the expansion assumes regular Green functions and a nonzero spectral denominator. It is a symmetry-allowed form, not a universal constitutive relation or a small-$Q$ approximation. Since $Q\to Q_N\neq0$, $\mathcal C_y$ scales as $1-T/T_N$, as does $P_y$ at generic fixed momenta with a nonzero quadratic coefficient. In contrast, $\Delta_{\rm eff}$ is linear in $|N|$ to leading order.

\subsection{Numerical test of the critical scaling}
\begin{figure}[t]
\centering\includegraphics[width=\columnwidth]{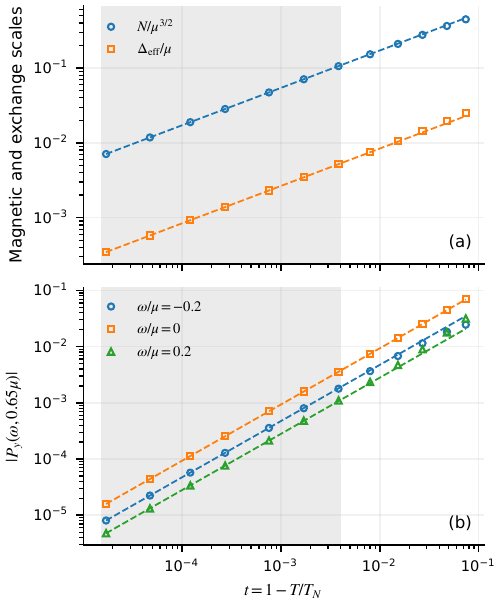}
\caption{Near-critical scaling from the coupled magnetic and infalling Dirac equations. (a) Condensate and static exchange; dashed lines have slope $1/2$. (b) Absolute spin polarization at fixed physical momentum $k/\mu=0.65$ and three frequencies; dashed lines have slope one. Reference lines are anchored at the closest computed temperature, not fitted. Shading identifies the seven-point fitting window $t<0.004$, with $t=1-T/T_N$.}
\label{fig:critical}
\end{figure}
We continue the circular branch with amplitude-normalized fields $\varphi=A_h z^{3/2}\chi$ and $a=A_h^2 B$, where $\chi$ equals one at the numerical horizon endpoint. Dividing the scalar, gauge and pitch equations by the corresponding powers of $A_h$ permits continuation from the normalized $A_h=0$ eigenproblem without a vanishing scalar boundary residual. The resulting $T_N/\mu=0.2879975926$ and $Q_N/\mu=0.8767431898$ agree with the independent linear calculation. At unchanged physical couplings we compute 12 ordered solutions with $1.71\times10^{-5}\leq t\leq0.0743$, and solve the Dirac equations on each profile. We fit $\ln |X|$ against $\ln t$ by unweighted least squares; $X$ is $N$, $\Delta_{\rm eff}$ or $P_y$.

Figure~\ref{fig:critical} shows distinct square-root and linear onsets. For the seven closest temperatures ($t<0.004$), the exponents of $N$, $\Delta_{\rm eff}$ and $P_y(0,0.65\mu)$ are $0.49966$, $0.50076$ and $1.00010$, respectively. Restricting to the five closest temperatures ($t<0.0008$) gives $0.49990$, $0.50023$ and $1.00003$; using all 12 gives $0.49639$, $0.50749$ and $1.00101$. At the 12 preselected points $\omega/\mu\in\{-0.2,0,0.2\}$ and $k/\mu\in\{0.35,0.65,1,1.3\}$, the seven-point polarization exponents span $0.9926$--$1.0057$. The signed coefficients vary across the cut; the exponent is not asserted at a node. The results thus test Eq.~\eqref{eq:scaling} with a dissipative self-energy. The same window gives exponents $1.00010$ for $Q_N-Q$ and $1.99884$ for $-\delta\Omega$, consistent with a smooth mean-field branch. As $Q_N$ is finite, these results are consistent with the vector-chirality scaling in Eq.~\eqref{eq:chirality}; they do not independently determine the small-$k$ coefficient in Eq.~\eqref{eq:chiralitypol}.

The two numerical runs differ in radial endpoints and integration tolerances (Appendix~\ref{app:num}); maximum relative differences over the 12 temperatures are $4.3\times10^{-5}$ for $\Delta_{\rm eff}$ and $2.2\times10^{-3}$ for $|P_y|$ at the 12 nonzero momenta--frequency points. Reducing the frequency regulator by ten at six temperatures changes these observables by at most $1.9\times10^{-5}$ and $6.7\times10^{-4}$, respectively. These are numerical sensitivity checks, not statistical uncertainties or tests of coupling universality.

\subsection{Numerical spectra and self-energy}
\begin{table}[t]
\centering
\caption{Computed condensate, pitch and static exchange. The onset row denotes the incipient ordering wavevector; the normal state itself has no physical pitch. Values are rounded.}
\label{tab:vals}
\begin{tabular}{rrrr}\toprule
$T/T_N$ & $N/\mu^{3/2}$ & $Q/\mu$ & $\Delta_{\rm eff}/\mu$\\\midrule
1.000 &0&0.8767&0\\
0.988&0.1873&0.8763&0.00935\\
0.926&0.4481&0.8737&0.02485\\
0.837&0.6303&0.8700&0.03977\\
0.649&0.8237&0.8621&0.06305\\
0.401&0.9157&0.8519&0.08079\\\bottomrule
\end{tabular}
\end{table}
\begin{figure}[t]
\centering\includegraphics[width=\columnwidth]{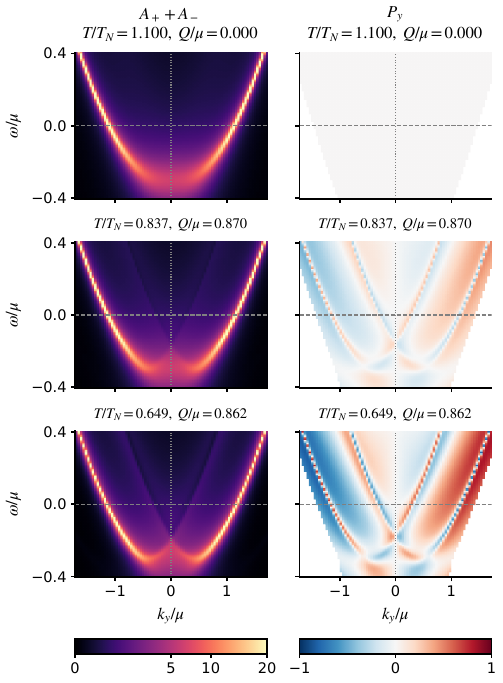}
\caption{Longitudinal laboratory-frame spectra above and below onset at fixed couplings. Left: $A_++A_-$. Right: $P_y$ shown only where the summed spectral weight exceeds 0.08 in the numerical normalization; masked pixels are not evidence for a physical node. Rows show the three temperatures from highest to lowest. Dashed horizontal lines indicate the Fermi level. The normal state uses $Q=0$ as a plotting convention. Odd parity and the node at $k=0$ follow from Eq.~\eqref{eq:odd}, independent of the display threshold.}
\label{fig:spec}
\end{figure}
Table~\ref{tab:vals} quantifies the distinct temperature dependences of the pitch and exchange. Between onset and $0.649T_N$, $Q$ decreases by $1.7\%$, while $\Delta_{\rm eff}$ grows from zero to $0.0631\mu$. At that temperature and $p=0$, the Hermitian mixing magnitudes at $\omega/\mu=-0.2,0,+0.2$ are $0.0445,0.0631,0.0707$ in units of $\mu$. The endpoint change is about $59\%$ relative to its negative-frequency value. This is a finite-window comparison, not a general monotonicity claim. A constant static exchange cannot represent the full mixing response over this window.

At the laboratory point $(\omega,k)=(0,0.65\mu)$ and $1.10T_N$, both spin channels have $\Sigma^R_{\rm eff}/\mu=-0.0672-0.0780i$. At $0.649T_N$, the corresponding values are $-0.0646-0.1210i$ and $-0.1689-0.1109i$ for $s=+$ and $-$. Thus both the energy shift and the dissipative self-energy are spin dependent; the half-difference is $(0.05215-0.00505i)\mu$. These imaginary parts characterize damping in the inverse Green function; they are not necessarily the fitted linewidths of quasiparticle peaks. Figure~\ref{fig:spec} shows reconstructed spectral weight and dispersive ridges through $\omega=0$. At finite temperature these are evidence for surviving low-energy spectral excitations, not a proof of a sharp zero-temperature Fermi surface or finite dc conductivity.

\section{Discussion and conclusions}
The central result is the connection between a stationary magnetic pitch and a dissipative, symmetry-constrained electronic response computed on the same magnetic solution. In contrast to a prescribed exchange field, both $Q(T)$ and $\Delta_{\rm eff}(T)$ emerge from the coupled radial equations and the probe fermion calculation at fixed model couplings. The exact parity identity and the numerically verified square-root versus linear onset distinguish exchange from the spectral polarization. The latter shares the symmetry and leading temperature scaling of vector chirality. Chirality-dependent odd-parity spin textures are known from microscopic spiral models \cite{larsen,gobel}; our scaling test concerns the dissipative spectral polarization at fixed frequency and momentum, not just the spin expectation of a clean band. For a fixed magnetic structure and a resolved domain, magnetic Bragg intensity is proportional to $N^2$ (after accounting for geometrical factors); a comparison with spin-resolved photoemission at a fixed regular $(\omega,k)$ could therefore test the predicted linear scaling with Bragg intensity. Opposite spiral domains can cancel the signed polarization. This is a prospective test, not a fit to EuSi data. The numerical values of $T_N$, $Q_N$ and the self-energy depend on the selected exchange kernel and other couplings. Appendix~\ref{app:lattice} provides an independent benchmark for the exchange mechanism and for the coexistence of a polarization node with clean gapless bands; it is not a second solution of the holographic equations.

EuSi exhibits sign-changing $S_y$ polarization of order five percent in selected near-Fermi-level features, commensurate spiral order and reconstruction separations exceeding 100 meV \cite{cheng}. The sign change and suppression near the node agree qualitatively with the computed longitudinal spin projection. Quantitative agreement in polarization magnitude is not claimed: the continuum spectra omit orbital-dependent photoemission matrix elements, domain populations and experimental resolution. The role of orbital character and the tendency to incommensuration have been discussed in Refs.~\cite{larsen,lee}. Our pitch is selected by an engineered exchange kernel rather than an itinerant microscopic instability and lacks crystalline lock-in; the lower magnetic transition in EuSi is outside our circular ansatz.

Matching $T_N$ alone to 62 K would set $\mu=k_B(62\,\mathrm K)/0.2880\simeq18.6$ meV and give $\Delta_{\rm eff}\simeq1.17$ meV at $0.649T_N$. Its twice-value, about 2.35 meV, is far smaller than the reported reconstruction separations. Since this static diagnostic is not a spectral peak separation, this is a scale warning rather than a quantitative discrepancy between identical observables. A material-specific test requires a common peak definition, orbital-resolved dispersion, crystalline pinning and resolution convolution. The separate lattice benchmark uses prescribed parameters in eV and cannot repair this scale mismatch.

The main limitations are the probe approximation, the phenomenological double-trace deformation, the restricted magnetic ansatz, the single coupling choice, and the longitudinal, helicity-averaged fermionic projection. The reported convergence checks concern numerical precision, not sensitivity to model assumptions. Future calculations should compare competing saddles and their fluctuation spectra, vary the deformation and hybridization, include gravitational backreaction, extend the spectral calculation to transverse momenta and compute the current correlator. Within its stated regime, the model demonstrates how finite-density holography can generate spontaneous spatial spin order and an odd-parity, frequency-dependent electronic self-energy without inserting a temperature-dependent exchange by hand.

\appendix
\section{Exact lattice benchmark for circular exchange}\label{app:lattice}
As a check complementary to microscopic spiral models \cite{brekke,gobel}, consider $V(y)=D(\tau_x\cos Qy+\tau_z\sin Qy)$ and a spin-independent bare (or bath-dressed) propagator $g_0$, the Fourier components $V_{\pm Q}=D(\tau_x\mp i\tau_z)/2$ satisfy $V_{sQ}V_{-sQ}=D^2P_s$. Each spin-momentum state couples to only one opposite-spin partner, giving the exact diagonal Green function
\begin{equation}
g_s(\omega,\boldsymbol{k})=\left[g_0^{-1}(\omega,\boldsymbol{k})-D^2g_0(\omega,\boldsymbol{k}-sQ\hat{\boldsymbol y})\right]^{-1}.
\label{eq:lattice}
\end{equation}
The spin-odd exchange self-energy is $D^2[g_0(\boldsymbol{k}-Q\hat{\boldsymbol y})-g_0(\boldsymbol{k}+Q\hat{\boldsymbol y})]/2$ at fixed frequency. It changes sign with handedness and vanishes on the mirror plane of a mirror-even band. For small $Q$ and a smooth $g_0$ it is $-D^2Q\partial_{k_y}g_0+O(D^2Q^3)$, giving a term proportional to $D^2Qk_y$ near the node if the second derivative is nonzero. No small-$Q$ assumption is required for Eq.~\eqref{eq:lattice}. In the clean limit the reconstructed eigenvalues in a block are
\begin{align}
E_{\pm,s}(\boldsymbol{k})&=\tfrac12[\epsilon_{\boldsymbol{k}}+\epsilon_{\boldsymbol{k}-sQ\hat{\boldsymbol y}}]\nonumber\\
&\quad\pm\sqrt{D^2+\tfrac14[\epsilon_{\boldsymbol{k}}-\epsilon_{\boldsymbol{k}-sQ\hat{\boldsymbol y}}]^2}.
\label{eq:bands}
\end{align}
At a bare degeneracy the splitting is $2|D|$; a global gap is not implied.

For the numerical benchmark we set $\epsilon_{\boldsymbol k}=-2\sum_i t_i\cos k_i-\mu_b$, with $(t_x,t_y,t_z)=(0.12,0.18,0.06)$ eV, $\mu_b=-0.30$ eV, $Q=\pi/3$ per effective layer and $D=0.10\sqrt{\max(0,1-T/(62\,\mathrm K))}$ eV. This $D(T)$ is prescribed, not the holographic $\Delta_{\rm eff}$. The bath in $g_0^{-1}=\omega-\epsilon-\Sigma_b$ uses an AdS$_2$-inspired thermal frequency dependence \cite{faulkner}:
\begin{equation}
\Sigma_b=-i\gamma_0-ic E_*^{1-2\nu}(2\pi k_BT)^{2\nu}
\frac{\Gamma(1/2+\nu-i\omega/(2\pi k_BT))}{\Gamma(1/2-\nu-i\omega/(2\pi k_BT))},
\label{eq:bath}
\end{equation}
with $\nu=0.35$, $c=0.30$, $E_*=0.20$ eV and $\gamma_0=2$ meV. At $k_y=Q/2$, $k_x=k_z=\pi/2$, the two $+$-channel spectral maxima are separated by 29.2, 68.4 and 88.3 meV at 58.9, 46.5 and 37.2 K, respectively. The clean splittings are 44.7, 100.0 and 126.5 meV, respectively. These are two peaks of one spin channel, not a direct difference between opposite spin spectra. The analytical Green function agrees with a direct six-layer folded-matrix inversion to $7\times10^{-15}$ in a spot check.
\begin{figure}[t]
\centering\includegraphics[width=\columnwidth]{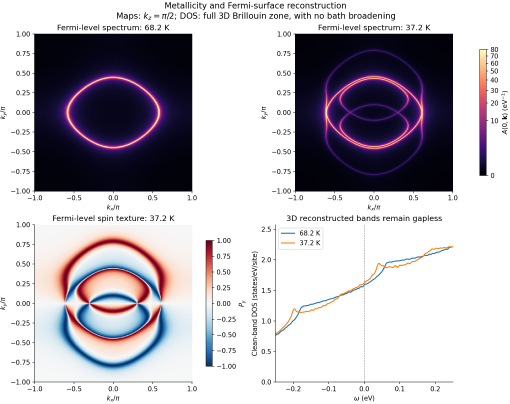}
\caption{Independent lattice benchmark with prescribed $D(T)$ and fixed $Q$. Top: Fermi-level spectral contours at $k_z=\pi/2$. Bottom left: bath-dressed polarization. Bottom right: clean-band density of states smoothed by 3 meV for display. These are model calculations, not EuSi data or holographic spectral grids.}
\label{fig:lattice}
\end{figure}
The clean reconstructed bands cross zero energy. Counting states within a 10 meV window about $E_F$ on a $128^3$ grid gives estimated densities of 1.59 and 1.63 states/(eV site) at 68.2 and 37.2 K, respectively (Fig.~\ref{fig:lattice}). These estimates use clean bands rather than dissipative tails. They establish metallic bands in this benchmark but do not establish transport in the holographic model.

\section{Variational and numerical prescription}\label{app:num}
\subsection{Magnetic free energy and boundary conditions}
At fixed $T,\mu$, the dimensionless horizon-unit functional (with the normal magnetic action subtracted) is
\begin{align}
\mathcal I=&\lim_{\varepsilon\to0}\left\{\int_\varepsilon^1 dz\left[\frac{f\varphi'^2+(\hat q-a)^2\varphi^2}{2z^3}
+\frac{f a'^2}{2g_s^2z}+\frac{V(\varphi)}{z^5}\right]\right.\nonumber\\
&\left.+\frac{3\sqrt{f(\varepsilon)}}{4\varepsilon^4}\varphi(\varepsilon)^2\right\}+\frac12 z_h\kappa(Q)\alpha^2.
\label{eq:func}
\end{align}
Here $\delta\Omega/\mu^4=z_h^{-4}\mathcal I$ per $C_N$. The scalar counterterm follows the holographic renormalization prescription \cite{skenderis} and cancels the leading divergence, and the renormalized variation at fixed $Q$ is $(-\beta+z_h\kappa\alpha)\delta\alpha$. The gauge source is held at zero. Varying $Q$ at fixed $T,\mu$ on the source-free saddle yields Eq.~\eqref{eq:pitch}. Multiplying Eq.~\eqref{eq:phi} by $\varphi$ and integrating by parts replaces the scalar quadratic and kinetic terms by a boundary term and $-u\varphi^4/(4z^5)$. The finite scalar boundary contribution $-\alpha\beta/2$ cancels the deformation term using Eq.~\eqref{eq:bc}, leaving
\begin{equation}
\frac{\delta\Omega}{\mu^4}=z_h^{-4}\int_0^1 dz\left[\frac{f(a')^2}{2g_s^2z}-\frac{u\varphi^4}{4z^5}\right].
\label{eq:omega}
\end{equation}
Both expressions use the same normalization and fixed-$\mu$ ensemble. Horizon regularity requires
\begin{align}
f'(1)\varphi'(1)&=[(\hat q-a_h)^2+m^2+u\varphi_h^2]\varphi_h,\nonumber\\
f'(1)a'(1)&=-g_s^2(\hat q-a_h)\varphi_h^2.
\end{align}
The nonlinear results use collocation for $w=\varphi/z^{3/2}$ and $a$, with $Q,b$ as free parameters; continuation is in the rescaled scalar at the near-horizon endpoint. The integration interval is $[10^{-5},1-2\times10^{-5}]$. The reported refinement (halving both cutoff distances and tightening tolerance) changes $T,Q,N$ at $0.649T_N$ by relative amounts $6.5\times10^{-6},2.1\times10^{-7},3.3\times10^{-6}$, respectively.

For an independent linear-onset check, $a=0$ and $w$ satisfies
\begin{equation}
(fw')'+\left[\frac{3f'}{2z}+\frac{m^2(f-1)}{z^2}-\hat q^2\right]w=0.
\label{eq:linw}
\end{equation}
Set $w(1)=1$ and $w'(1)=(\hat q^2+m^2)/f'(1)-3/2$. Integrate inwards using a first-order horizon expansion, and extract $\alpha=w(0)$ and $\beta=w'(0)$ by UV Taylor expansion. A DOP853 integration with relative tolerance $10^{-10}$, absolute tolerance $10^{-12}$, and both cutoff distances $10^{-6}$, followed by a two-variable root solve for $r$ and $r_Q$ (centered difference with step $10^{-4}$), gives $b=0.5437747$, $T_N/\mu=0.2879976$, $Q_N/\mu=0.8767432$ and $r(T_N,0)=0.5443572$. A centered second difference with step $10^{-3}$ gives $r_{QQ}=5.7434$. A centered difference in $b$ with step $10^{-4}$ gives $\partial r/\partial(T/\mu)=0.5745$ at fixed $Q$, so lowering $T$ makes this eigenchannel unstable. Changing the cutoffs to $10^{-5}$ or $10^{-7}$ at this root changes the evaluated $r$ by less than $10^{-9}$. The independent shooting check concerns the normal-state onset. The additional amplitude-normalized continuation and Dirac runs directly test near-onset nonlinear and spectral scaling.

\subsection{Dirac prescription and numerical checks}
For each spacetime helicity $\ell=\pm1$, propagate two independent infalling flavor solutions assembled into matrices $U,V$ after removal of the spacetime spin connection. With $W=z_h[\omega+q_\Psi(1-z^2)]$, the radial equations in the chosen gamma-matrix basis are \cite{faulkner}
\begin{align}
U'&=\frac{M}{z\sqrt f}U+i\left(\frac{W}{f}-\frac{\ell K}{\sqrt f}\right)V,\label{eq:u}\\
V'&=-\frac{M}{z\sqrt f}V+i\left(\frac{W}{f}+\frac{\ell K}{\sqrt f}\right)U.\label{eq:v}
\end{align}
For the $e^{-i\omega t}$ convention, the infalling subspace has $U=V$ at the horizon up to the common factor $(1-z)^{-i\omega/(4\pi T)}$, with $\omega,T$ in the same units. The massless UV prescription is $G^R_\ell=-iUV^{-1}$ at the boundary, and $G^R=(G^R_++G^R_-)/2$. Integration in $\log[z/(1-z)]$ runs from $z=1-10^{-8}$ to $10^{-5}$ with regulator $\omega\to\omega+2\times10^{-5}i$ in scaled units. The reported cutoff and tolerance refinement changes the sampled response by $2.8\times10^{-5}$ in the maximum norm normalized to the maximum sampled response. Spin-resolved spectral weights are nonnegative on the computed mesh; tested parity residuals and nodal polarizations are below $1.4\times10^{-11}$. These checks do not quantify systematic uncertainty under parameter variation or establish positivity at all complex frequencies.

\subsection{Near-critical continuation and precision}
For the additional critical calculation we set $A_h=\varphi(z_H)/z_H^{3/2}$, normalize $\chi(z_H)=1$ and divide the accumulated pitch integral by $A_h^2$. The normalized equations retain the same UV and first-order horizon boundary expansions. We compute the $A_h=0$ limiting solution separately for each cutoff choice to define its own $T_N$, avoiding a spurious temperature offset in the smallest $t$. The baseline uses $(z_{\rm UV},1-z_H)=(10^{-5},2\times10^{-5})$ and collocation tolerance $10^{-8}$. The cutoff-refined run uses $(5\times10^{-6},10^{-5})$ and tolerance $3\times10^{-8}$; its looser collocation tolerance avoids residual stagnation at the singular endpoints. The magnetic residual is below $3\times10^{-8}$ in this run. For the two Dirac runs the relative tolerances are $10^{-8}$ and $2\times10^{-9}$, and horizon cutoff distances $10^{-8}$ and $2.5\times10^{-9}$, with absolute tolerances $0.05$ times the relative tolerance. Both initially use $\eta/\mu=2\times10^{-5}$; a six-temperature run uses $2\times10^{-6}$. The refined zero-amplitude calculation has $|P_y|<3\times10^{-15}$ even with an arbitrary finite rotating-frame pitch. Across the new temperatures, the parity residual at $(0,\pm0.65\mu)$ is below $6\times10^{-13}$ and the nodal polarization at $(0,0)$ is below $2\times10^{-13}$.

\section*{Data and code availability}
The accompanying source package includes the original magnetic and fermionic solvers and archived profiles, free-energy checks, spectral grids and figure scripts, as well as the independent onset check, normalized near-critical continuation, raw numerical outputs and fitting/plotting script. A public archive identifier can be added upon deposition.

\end{document}